\documentstyle[11pt,newpasp,twoside,epsf]{article}
\def\edcomment#1{\iffalse\marginpar{\raggedright\sl#1\/}\else\relax\fi}
\reversemarginpar

\begin{document}
\title{Central DM density cuspiness in LSB's: a stellar
kinematics approach.}
 \author{A. Pizzella, E.M. Corsini, F. Bertola, and L. Coccato}
\affil{Dipartimento di Astronomia, Universit\'a di Padova, Italy}

\author{J. Magorrian, and M. Sarzi}
\affil{Physics Department, University of Oxford, UK}

\author{J.G. Funes, S.J.}
\affil{Vatican Observatory, Tucson, USA}

\begin{abstract}
We present preliminary results from spectroscopic observations of a
sample of 11 low surface brightness galaxies (LSB). We measured the
stellar and gaseous kinematics along their major and minor axes. Such
information will allow us to accurately investigate the dark matter
(DM) content within their optical regions, providing further
constraints on 
the predictions of standard CDM models. Although dynamical modeling is
still in progress, our observations already show that the derived
stellar kinematics is more regular than the ionized gas one, which
often shows evidence for non-circular and asymmetric motions.
\end{abstract}

\section{Observations and results}
We obtained long-slit spectroscopic data with the VLT FORS2
spectrograph ($\sigma_{instrument}=45\,km/s$) in the H$_\beta$,
[OIII]5007\AA, MgII and Fe region for 11 LSBs with the aim of
constructing dynamical mass models based on both the stellar and
ionized gas kinematics to constrain their DM density radial profiles,
in particular in their nuclear regions. This part of the work is still
in progress.
However, we should already stress the following two important
observational results: {\bf 1)} The ionized gas is subject to
non-circular motion in the majority of our sample galaxies. {\bf 2)}
The stellar kinematics is generally very regular. The high degree of
symmetry of the stellar velocity and velocity dispersion curves and
the small scatter between the points allows to derive the stellar
kinematical profiles with small error-bars and without applying any
smoothing to the data. This is not always true
for the gas kinematics. Such precise kinematical measurement are
therefore very encouraging for the forthcoming dynamical mass
models. In Fig.1a we plot the velocity and velocity dispersion radial
profiles of ESO 189-07: {\it Left} and {\it right panels} represent
the stellar and ionized gas kinematics respectively.  The {\it lower}
and {\it upper panels} represent the major and minor axis
respectively. The curves have been folded around the kinematical
center and the two symbols refer to the two side of the galaxy (with
the exception of the gaseous minor axis which has not been
folded). Deprojected velocities and radii in Kpc are also shown. The
remarkable degree of symmetry of the stellar kinematics also allows to
check the positioning of the slit and an accurate determination of the
kinematical center. Conversely, the ionized gas kinematics is much
less defined and such operations would be less reliable if based on
it.
\begin{figure}
\plottwo{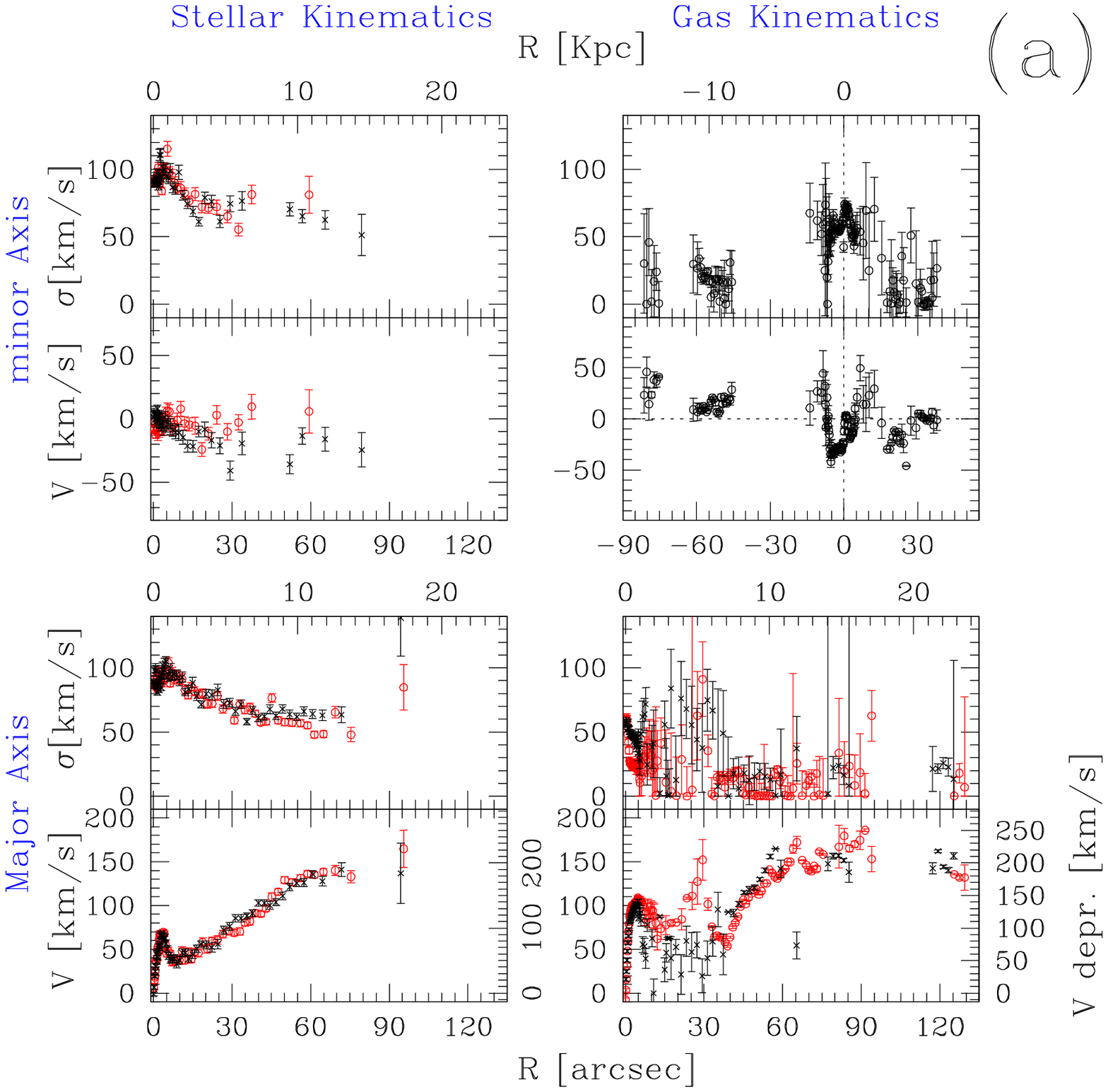}{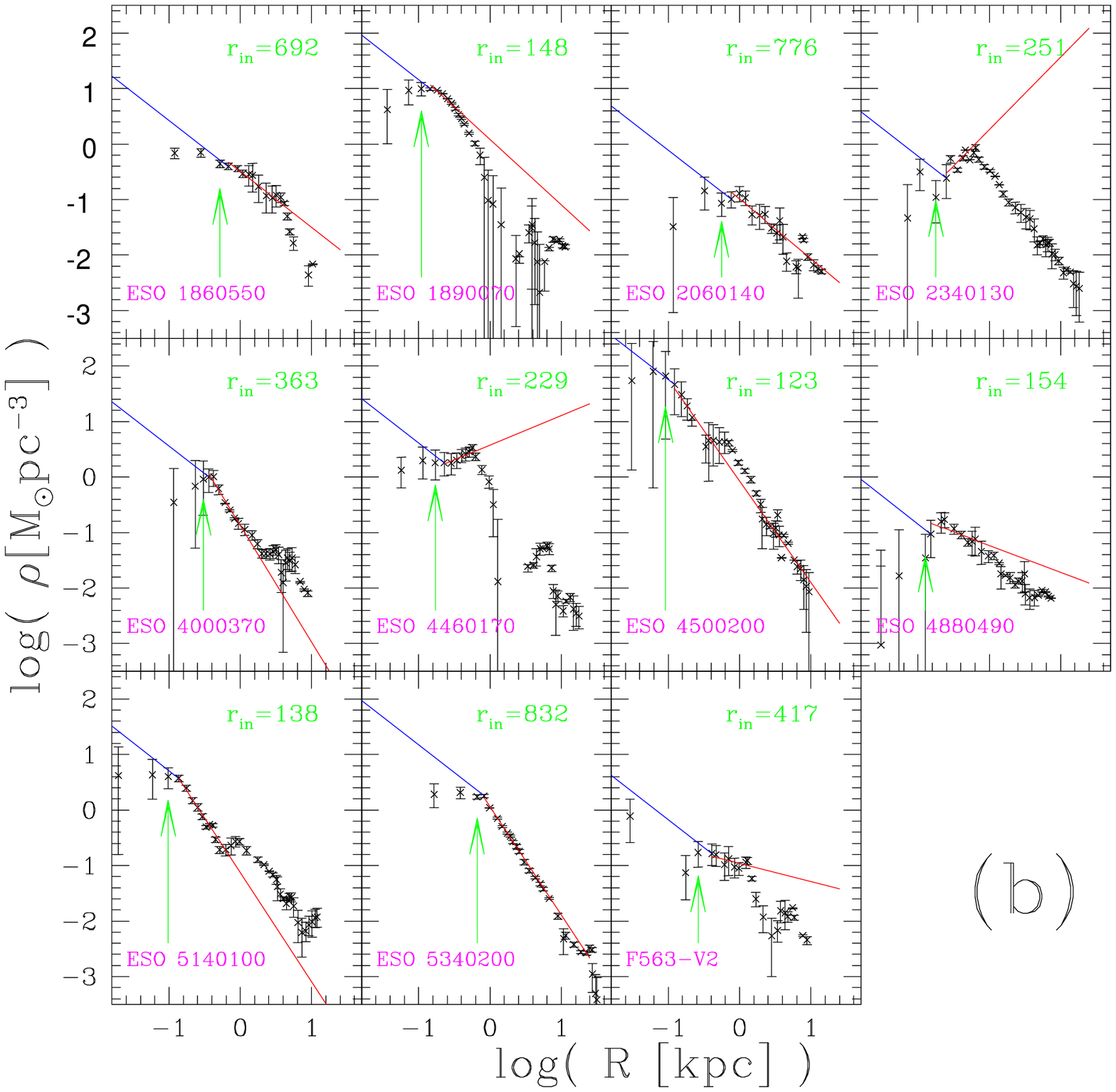}
\caption{{\it(a)} Velocity and velocity dispersion radial profiles of ESO 189-07.
{\it(b)} Radial density profile derived for the 11 sample
galaxies. See text for a full description of the figures.  }
\end{figure}
Following the scheme indicated by previous authors (de Blok et al
2001, Swaters et al. 2003) we derived the central mass density radial
profile directly from the major axis ionized gas velocity curve with
the only assumption of spherical mass distribution.  We took the major
axis velocities, deprojected, smoothed and derived the density $\rho
(r)$. The resulting radial density profiles are plotted in Fig.1b. In
the figure we show the value of $r_{in}$ (in parsec) defined as the
innermost radius at which we can derive a reliable density value
(i.e. the velocity measurement in not affected by seeing smearing or
pixel size). The inner line is the density profile $\rho\propto
r^{-1}$ passing through $r_{in}$. We also show as a more external line
the density $\rho\propto r^\alpha$ fit in the external regions
adiacent to $r_{in}$ as comparison. From the figure it is possible to
see that we find both galaxies showing a density radial profile less
peaked or as peaked as $\rho\propto r^{-1}$ in the center.  It is also
evident that there are cases as ESO4460170 or ESO2340130 where the
density decreases inward in the center. This is an artifact due to the
presence of ionized-gas non-circular motions in the core region,
motions that we can not see from the major axis alone but we know are
present since we have minor axis kinematical information.  But is the
gas the best tracer of the potential in the core region ?

\end{document}